\documentclass[aps,prb,twocolumn,showpacs,showkeys,footinbib,superscriptaddress]{revtex4-2}%superscriptaddress

\usepackage{amsmath}
\usepackage{amssymb}
\usepackage{graphicx}
\usepackage{bm}
\usepackage{color}
\usepackage{relsize}
\usepackage{braket}
\usepackage[caption=false]{subfig}

\usepackage{hyperref}
\usepackage[all]{hypcap}

\usepackage{slashbox}

\begin{document}
\title{ Effects of dynamical dielectric screening on the excitonic spectrum of monolayer semiconductors }
\date{\today}

\author{Dinh Van Tuan}
\email[]{vdinh@ur.rochester.edu}
\affiliation{Department of Electrical and Computer Engineering, University of Rochester, Rochester, New York 14627, USA}
\author{Hanan~Dery}
\email[]{hanan.dery@rochester.edu}
\affiliation{Department of Electrical and Computer Engineering, University of Rochester, Rochester, New York 14627, USA}
\affiliation{Department of Physics and Astronomy, University of Rochester, Rochester, New York 14627, USA}

\begin{abstract} 
We  present a new method to solve the dynamical Bethe-Salpeter Equation numerically. The  method allows one to investigate  the effects of  dynamical dielectric screening on the spectral position of excitons in  transition-metal dichalcogenide monolayers. The dynamics accounts for the response of optical phonons in the materials below and on top the monolayer to the electric field lines between the electron and hole of the exciton.    The inclusion of this effect unravels the origin of a counterintuitive energy blueshift of the exciton resonance, observed recently in monolayer semiconductors that  are  supported on ionic crystals with large dielectric constants. A surprising result is that while energy renormalization of a free electron in the conduction band or a free hole in the valence band  is controlled by the low-frequency dielectric constant, the bandgap energy introduces a phase  between the photoexcited electron and hole, rendering contributions from the high-frequency dielectric constant also important when evaluating self-energies of the exciton components. As a result, bandgap renormalization of the exciton is not the sum of independent contributions from energy shifts of the conduction and valence bands. The theory correctly predicts  the  energy shifts of exciton resonances  in various  dielectric environments that embed two-dimensional  semiconductors.
\end{abstract}


\maketitle

%%%%%%%%%%%%%%%%%%%%%%%%%%%%%%%%%%%%%%%%%%%%%%%
%%%%%%%%%%%%%%%%%%%%%%%%%%%%%%%%%%%%%%%%%%%%%%%
%%%%%%%%%%%%%%%%%%%%%%%%%%%%%%%%%%%%%%%%%%%%%%%
%%%%%%%%%%%%%%%%%%%%%%%%%%%%%%%%%%%%%%%%%%%%%%%
\section{Introduction}

Transition-metal dichalcogenides (TMDs) have emerged as an excellent platform to study various exotic quantum effects ranging from  valley-dependent selection rules in the monolayer  limit   \cite{Xiao_PRL12,Mak_PRL10,Splendiani_NanoLett10}   to fractionally quantized anomalous Hall effect recently observed in twisted MoTe$_2$ bilayers \cite{Cai_Nat_23,Park_Nat_23,Xu_PRX_2023,Goldman_PRL_2023,Dong_PRL_2023}.    Various interesting properties of these materials originate from strong  spin-orbit  and  Coulomb interactions which support the formation of various excitonic states, such as neutral and charged excitons \cite{Mak_NatMater13, Chernikov_PRL14,He_PRL14,Stier_PRL2018,Liu_PRB2019,Jones_NatPhys16,Plechinger_NatComm2018,Plechinger_NanoLett_2016,Wang_NanoLett17,Wang_NatNano2017,Courtade_PRB17}, neutral and charged biexcitons \cite{Hao_NatComm2017,Chen_NatComm2018,Ye_NatComm2018,Li_NatComm2018,Barbone_NatComm2018}, as well as hexcitons (six-body states) and oxcitons (eight-body states)  \cite{VanTuan_PRL22,VanTuan_PRB22,Choi_arXiv2023}. The two-dimensional nature of TMD monolayers makes them strongly susceptible to  changes in the dielectric environment around them.   Raja \textit{et al.} showed  that the bandgap and exciton binding energies of TMD monolayers can change  by hundreds of meV  using different dielectric environments \cite{Raja_NatComm17}. Their study further pointed out that the offset between strong changes in bandgap and binding energies leads to small change in the spectral position of the neutral exciton.    An overall small energy redshift was observed in the spectrum of devices in which the effective dielectric constant of the surrounding environment was larger \cite{Raja_NatNano19}. 

However, a recent experiment using  titanium-based oxides to support TMD monolayers  showed an opposite trend \cite{Amine_XYZ24}. The optical spectra  were measured in hBN-encapsulated TMD devices  and then in devices  where  the supporting hBN layer  was replaced by  titanium dioxide (TiO$_2$) or  strontium titanium oxide (STO - SrTiO$_3$).  This replacement led to energy blueshift of the exciton  resonance in spite of the larger dielectric constants of titanium-based oxides, where  the energy blueshift seemed to be commensurate  with the amplitude of the dielectric constant. The apparent contradiction of these   findings with previous results   clearly suggests that  understanding of the physics is lacking.

 Here, we tackle  the problem by developing a new method to solve the Bethe-Salpeter Equation (BSE) in the dynamical regime. We use  a frequency-dependent  dielectric function through the response of optical phonons  in the surrounding materials to the electric field induced by the  interaction between the electron and hole of the exciton.  The results show that band gap renormalization (BGR) of single particles - free electrons (holes) in the conduction (valence) band -  is governed by the low-frequency dielectric constant, $\epsilon_0$. On the other hand, the bandgap energy renders the high-frequency dielectric constant, $\epsilon_\infty$, suitable in self-energy calculations of  the exciton's electron and  hole.   Furthermore, the low-frequency dielectric constants of  ionic crystals are proven to give  dominant contributions to the exciton binding energy.  The theory helps to explain measurements of TMD monolayers that are supported on TiO$_2$ and STO \cite{Amine_XYZ24}.  Since self-energies and binding energies have important contributions from $\epsilon_\infty$ and $\epsilon_0$, respectively, replacing the surrounding dielectric environment  can lead to energy blueshift of the exciton resonance  if the replacement introduces a small change in $\epsilon_\infty$ and a large change in  $\epsilon_0$ ($\Delta  \epsilon_\infty \ll \Delta  \epsilon_0$).

%%%%%%%%%%%%%%%%%%%%%%%%%%%%%%%%%%%%%%%%%%%%%%%
%%%%%%%%%%%%%%%%%%%%%%%%%%%%%%%%%%%%%%%%%%%%%%%
%\subsection*{Organization of the paper}

This paper is organized as follows. Section~\ref{sec:Mov} includes a brief derivation, still without dynamical effects,  that explains why the exciton resonance is expected to redshift in energy when the dielectric constant  increases.   To motivate the inclusion of  dynamical effects, we discuss the phonon spectrum of the dielectric materials used in the experiment and compare the phonon energies with  kinetic energies of the electron and hole in the exciton.  The theoretical formalism  is presented in Sec.~\ref{sec:theory}, where we explain how to include the dynamical dielectric function in self-energy calculations of  charge particles as well as  in the BSE. Simulation results and their analyses are presented in Sec.~\ref{sec:results}. A summary and outlook are given in Sec.~\ref{sec:con}. The appendices include the functional form of the dynamical potential,  technical details of the iterative method, Pad\'{e} approximation, parameters used in simulations,  and computation aspects.

%%%%%%%%%%%%%%%%%%%%%%%%%%%%%%%%%%%%%%%%%%%%%%%
%%%%%%%%%%%%%%%%%%%%%%%%%%%%%%%%%%%%%%%%%%%%%%%
%%%%%%%%%%%%%%%%%%%%%%%%%%%%%%%%%%%%%%%%%%%%%%%
%%%%%%%%%%%%%%%%%%%%%%%%%%%%%%%%%%%%%%%%%%%%%%% 
%%%%%%%%%%%%%%%%%%%%%%%%%%%%%%%%%%%%%%%%%%%%%%%
%%%%%%%%%%%%%%%%%%%%%%%%%%%%%%%%%%%%%%%%%%%%%%%
%%%%%%%%%%%%%%%%%%%%%%%%%%%%%%%%%%%%%%%%%%%%%%%
%%%%%%%%%%%%%%%%%%%%%%%%%%%%%%%%%%%%%%%%%%%%%%%
\section{Background and motivation}\label{sec:Mov}

We start by analyzing what happens to the exciton resonance energy when the dielectric screening of the environment increases. This brief analysis will motivate the need to incorporate the dynamical dielectric response of the encapsulating materials.

The energy shift of the exciton resonance is determined by the BGR and binding energy. The BGR of a semiconductor is attributed to the Coulomb-hole effect \cite{Haug_SchmittRink_PQE84}, reflecting the change in energy needed to excite an electron across the band gap when the Coulomb potential at the immediate vicinity of the electron is changed. In case that the dielectric environment is changed, one gets \cite{Haug_SchmittRink_PQE84,Scharf_JPCM19,Marauhn_PRB23}  
 \begin{eqnarray}
\Delta E_g  &=&   E_{g,2} - E_{g,1} = \sum_{{\bf q}} \left[V(\epsilon_2, {\bf q}) - V(\epsilon_1, {\bf q}) \right]  \nonumber \\
&=&  \lim_{r \rightarrow 0} \left[V(\epsilon_2, {\bf r}) - V(\epsilon_1, {\bf r})\right],
 \label{eq:eg}
\end{eqnarray}
where $\bf q$ is the transferred momentum in the Coulomb interaction and $\epsilon_{1,2}$ are effective  dielectric constants of the two environments. Considering first a small change in the dielectric constant, such that $\epsilon_2 = \epsilon_1  +\delta \epsilon $ where $\delta \epsilon$ is small and positive, we get 
\begin{equation}
\Delta E_g =  \delta \epsilon \,\,  V' (\epsilon_1, r = 0 ) 
\end{equation}
with  $V'(\epsilon_1, r)  = \partial V / \partial \epsilon $ for $\epsilon=\epsilon_1$.

Because the exciton state is calculated with the same Coulomb potential, we can use perturbation theory to quantify the change in exciton binding energy
\begin{equation}
\Delta E_b = \delta \epsilon \langle\,\,  \psi_1 | V' (\epsilon_1, r ) | \psi_1 \rangle   ,
\end{equation}
where $|\psi_1\rangle$ is the exciton wave function corresponding to $V(\epsilon_1, r)$. The total energy shift of the exciton resonance, $\Delta E_{X_0} = \Delta E_g - \Delta E_b $,  becomes
\begin{equation}
\Delta E_{X_0} =  \delta \epsilon \left\langle\,\,  \psi_1 \left| { \left[ V' (\epsilon_1, r = 0 ) -  V' (\epsilon_1, r ) \right]  } \right| \psi_1 \right\rangle.  
\end{equation}
Assuming small size exciton, the term in square brackets is merely the derivative of $r$, and we get 
\begin{equation} 
\Delta E_{X_0} \simeq   -\delta \epsilon \left\langle\,\,  \psi_1 \left|  ({\partial^2 V}/{\partial \epsilon  \partial r}) \times r  \right| \psi_1 \right\rangle  < 0 . \label{eq:red}
\end{equation}
The energy change is negative because $V(\epsilon, r)$ is a decaying function of both $\epsilon$ and $r$, so that  $\partial^2 V/ \partial \epsilon  \partial r >0$. In other words, the exciton energy \textit{redshifts} under a small increase  of the dielectric constant. This outcome is expected since the difference between the Coulomb potentials of two dielectric environments is largest at $r = 0$ from which the BGR is evaluated. Thus,  the energy redshift induced by BGR cannot be overshadowed by the blueshift induced by weaker binding energy.  We can further generalize the result for cases in which there is a large difference between $\epsilon_2$ and $\epsilon_1$, replacing the term $\delta \epsilon$ with an integration over $ \epsilon$ between $\epsilon_1$ and $\epsilon_2$.

From the above derivation one might expect the energy redshift to be a  prevalent effect whenever the effective dielectric screening increases. However, newly found experimental results show the exact opposite \cite{Amine_XYZ24}.  Using device structures in which TMD monolayers are covered by hBN and supported on STO,  the exciton resonance energy showed a  blueshift of $\sim$30~meV  with respect to hBN-encapsulated monolayer devices. Similar devices that are supported on rutile (TiO$_2$) showed an energy blueshift of $\sim$15~meV. These results are especially surprising given that the static dielectric constant of STO is greater than $10^4$ at low temperatures \cite{Sawaguchi_JPSJ62,Neville_JAP72} and that of rutile is greater than 10$^2$ \cite{Parker_PR61}.  Namely, instead of showing the strongest energy redshift in monolayers that are supported on STO, the measurements yielded the strongest energy blueshift. And instead of yielding the second strongest energy redshift in monolayers that are supported on rutile, the measurements yielded the second strongest energy blueshift. 

To understand the problem, we examine the state distributions of the exciton's electron and hole components due to their relative motion. The  stochastic variational method (SVM) \cite{VanTuan_PRB18,VargaBook,Varga_PRC95} is employed to find the exciton wavefunction in WSe$_2$ monolayer covered by hBN and supported on STO (hBN-WSe$_2$-STO). Solid lines show calculation results when using the low-frequency dielectric constants of hBN and STO, and dashed lines show the corresponding results when using their high-frequency dielectric constants. Materials parameters  are listed in Appendix \ref{app:Para}. Figure~\ref{fig:Phonon}(a) shows state distributions of the exciton components.  The distributions suggest that the kinetic energies from relative motions of the electron and hole in the TMD monolayer are in the energy range of optical phonons in the supporting STO, as shown in Fig.~\ref{fig:Phonon}(b). Incorporating the lattice dynamics of encapsulating ionic layers is important because the relative motion between the electron and hole means that the electric field lines that spread out of the monolayer, as shown in  Fig.~\ref{fig:Phonon}(c), change over time. 

%%%%%%%%%%%%%%%%%%%%%%%%%%%%%%%%%%%%%%%%%%%%%%%%%%%%%%%%%%
%%%%%%%%%%%%%%%%%%%%%%%%%%%%%%%%%%%%%%%%%%%%%%%%%%%%%%%%%%
%%%%%%%%%%%%%%%%%%%%%%%%%%%%%%%%%%%%%%%%%%%%%%%%%%%%%%%%%%
%%%%%%%%%%%%%%%%%%%%%%%%%%%%%%%%%%%%%%%%%%%%%%%%%%%%%%%%%%
%%%%%%%%%%%%%%%%%%%%%%%%%%%%%%%%%%%%%%%%%%%%%%%%%%%%%%%%%%
\begin{figure}[t] 
\centering
\includegraphics[width=8.5cm]{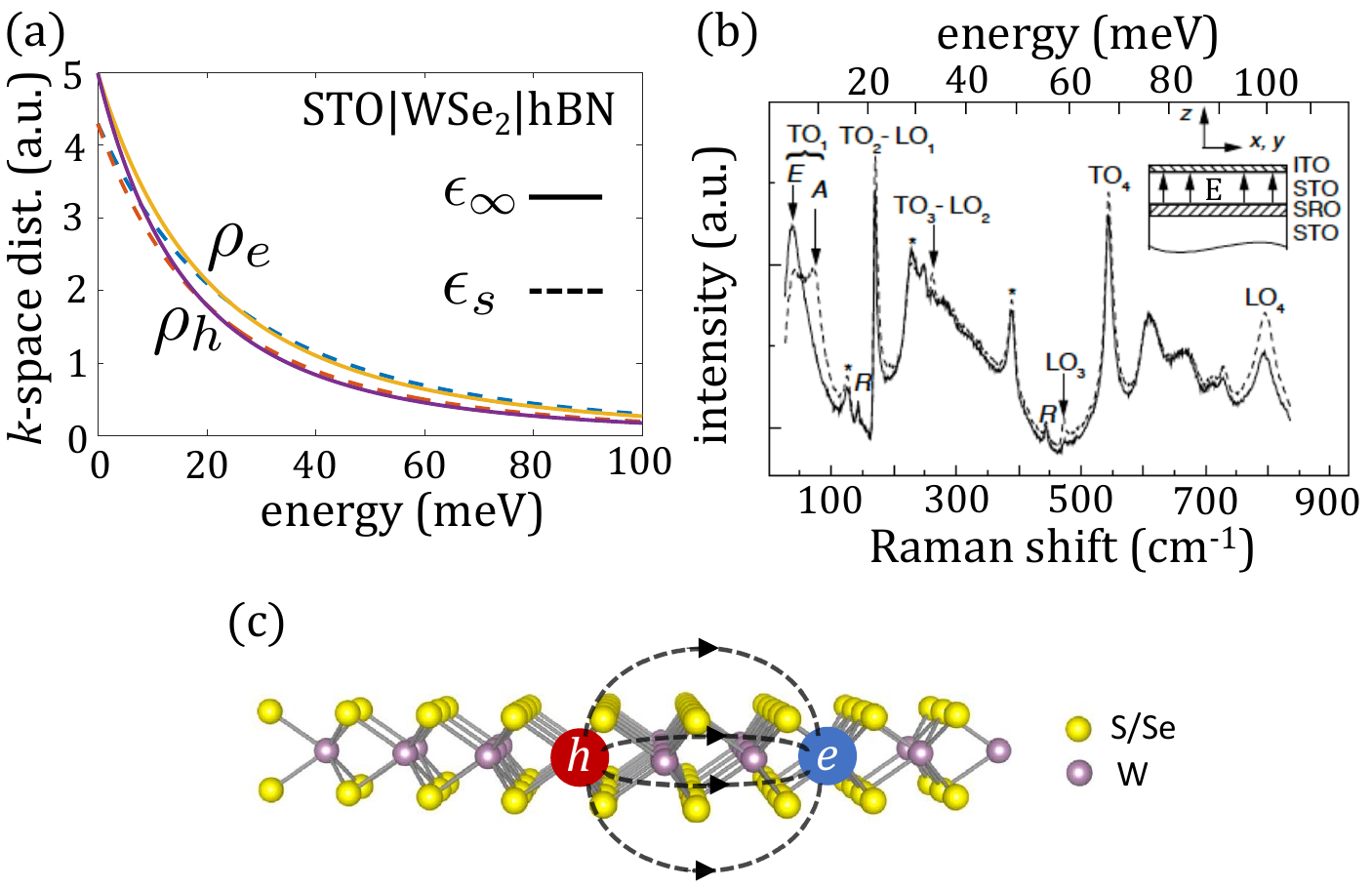}
\caption{ (a) k-space distributions of the electron and hole due to their relative motion in a charge-neutral exciton.  (b) The optical phonon modes in STO, extracted from Raman spectra of 1 mm STO film at $T=5$  K without electric field (solid line) and when a field $E = 2.2  \times 10^5$ V/cm (dashed lines) is directed normal to the film plane. Taken from Ref. \cite{Akimov_PRL2000}. (c) Schematic of the electric field lines between  the electron and hole of an exciton (dashed lines), some of which pass through  the surrounding environments.   }\label{fig:Phonon} 
\end{figure}

\section{Theory}\label{sec:theory}
The dynamical effects are obtained by solving the  BSE through inclusion of  the dynamical Coulomb interaction between charge particles. We  first show how to incorporate  the dynamical dielectric function in the potential and the way to obtain the dynamical self-energy of each component  of the exciton.  Then we  use these quantities in  the dynamical BSE. 

\subsection{Dynamical dielectric function}
We start by  considering  the frequency dependence of the dielectric function of a general material  subjected to a time-dependent periodic electric field of frequency $\omega$. The structure geometry  will later be incorporated to obtain the effective dielectric function of the Coulomb  interaction between charged particles in the monolayer. 

Assuming a lossless Lorentzian oscillator model, the permittivity of ionic crystals under the effect of  a periodic electric field is given by
\begin{eqnarray}
\epsilon(\omega) = \epsilon_{\infty} \,\,\prod_j \frac{\omega_{j,\text{LO}}^2 - \omega^2}{\omega_{j,\text{TO}}^2 - \omega^2}. 
\label{eq:LST}
\end{eqnarray}
The ratio between the static and high-frequency permittivities is the celebrated Lyddane–Sachs–Teller relation $\epsilon(\omega=0)/\epsilon(\omega=\infty) \equiv  \epsilon_0/\epsilon_\infty =  \prod_j \omega_{j,\text{LO}}^2 /\omega_{j,\text{TO}}^2 $. The index $j$ runs over the optical-phonon modes, where $\omega_{j,\text{LO/TO}}$ is the associated frequency of the longitudinal/transverse optical lattice vibration.  Materials like TiO$_2$ and STO have several longitudinal and  transverse modes such that $\omega_{j,\text{LO}} \simeq \omega_{j+1,\text{TO}}$. Figure~\ref{fig:Phonon}(b) shows the phonon frequencies of STO  
  with $\omega_{1,\text{LO}} = \omega_{2,\text{TO}}$, $\omega_{2,\text{LO}} = \omega_{3,\text{TO}}$, and $\omega_{3,\text{LO}} \simeq \omega_{4,\text{TO}}$. As a consequence, the frequency-dependent permittivity in Eq.~(\ref{eq:LST}) is mostly governed by the first and last modes, yielding

\begin{equation}
\epsilon(\omega ) = \epsilon_\infty  + \epsilon_\infty  \frac{\omega^2_{4,\text{LO}} - \omega^2_{1,\text{TO}} }{\omega^2_{1,\text{TO}} - \omega^2}.
\label{Eq:DynaDie}
 \end{equation}
%This functional form is also applicable for other materials like hBN in which  the  frequency-dependent permittivity is mainly governed by  two phonon modes \cite{Cai_SSC2007}.   
Based on this observation, we will use two unitless parameters to describe the dielectric response of each layer. The first parameter is the ratio between the first TO phonon energy and the estimated exciton binding energy 
\begin{equation} 
r_1 = \frac{\hbar \omega_{1,\text{TO}}}{\varepsilon_{\text{X}^0}}.
\end{equation}
We use $\varepsilon_{\text{X}^0} = 170$~meV in this work (approximately the exciton binding energy in hBN-encapsulated WSe$_2$ \cite{VanTuan_PRB18,Stier_PRL2018}).  The second parameter is the square frequency ratio between the last LO mode and first TO mode
\begin{equation} 
r_2 = \frac{\epsilon_0}{\epsilon_\infty} = \frac{\omega^2_{N,\text{LO}}}{\omega^2_{1,\text{TO}}}.  \label{Eq:r2}
\end{equation}
The dielectric function  in Eq.~(\ref{Eq:DynaDie}) can then be written as
\begin{equation} 
\epsilon(\omega) = \epsilon_\infty \left(  1 +  \frac{r_2 -1 }{1- \frac{\omega^2}{r^2_1 \varepsilon_{\text{X}^0}^2}} \right). 
\label{Eq:DynaDielec}
\end{equation} 

The Coulomb  interaction between charge particles  in the monolayer is obtained by solving the Poisson Equation with the appropriate structure geometry.  Details of the derivation are given in Ref.~\cite{VanTuan_PRB18}. The dynamical Coulomb potential between charge particles is given by 
\begin{equation}
V({\bf q}, \omega) = \frac{2 \pi e^2 }{ A  \,\,\epsilon({\bf q}, \omega) \,\,   q  }, 
\label{Eq:DynaPoten}
\end{equation}
where $A$ is the area of the system. Dynamical effects are incorporated  through the  dielectric functions of the surrounding dielectric materials, as detailed in  Appendix~\ref{ap:Ponten}.

The dynamical Coulomb potential  has singularities at the phonon frequencies. To circumvent this difficulty when solving the BSE or evaluating the self-energies, we use  finite-temperature Green's function formalism in which real frequencies are replaced by imaginary and discrete Matsubara frequencies \cite{MahanBook}. Namely, $\epsilon({\bf q}, \omega)$  is replaced with $\epsilon({\bf q}, z-z')$, where $z$ and $z'$ are imaginary Matsubara energies of fermions before and after the interaction. Their discretized energy form is $(2\ell+1)\pi i k_\text{B} T$, where $\ell$ is an integer and T is temperature. Consequently, the positive real number $\omega^2$ in Eq.~(\ref{Eq:DynaDielec}) is replaced by $(z-z')^2$ which is a  negative real number. Rather than having singularities, the dielectric function is now monotonously decaying from $\epsilon_0$ to $\epsilon_\infty$ as $z-z'$ departs from 0.   

%%%%%%%%%%%%%%%%%%%%%%%%%%%%%%%%%%%%%%%%%%%%%%%%%%%%%%%%%%
%%%%%%%%%%%%%%%%%%%%%%%%%%%%%%%%%%%%%%%%%%%%%%%%%%%%%%%%%%
\subsection{Dynamical Self-Energy}

The self-energy of an electron in  the conduction ($c$) or valence band (v) is calculated  from a self-consistent solution of the Dyson Equation
\begin{equation}
\Sigma_i({\bf k},z ) = - \frac{1}{\beta} \sum_{{\bf q},z'} G_i({\bf k- q}, z' ) V({\bf q}, z - z'). 
\label{eq:DynaSelfE}
\end{equation}
where  $\beta^{-1}=k_BT$, $i=\{\text{c,v}\}$, and the Green's function is
\begin{eqnarray}
G_i({\bf k},z ) &= &\frac{G_i^0({\bf k},z )}{1 - G_i^0({\bf k},z ) \Sigma_i({\bf k},z )}   \nonumber\\
&=& \frac{1}{z - \varepsilon_i({\bf k}) + \mu- \Sigma_i({\bf k},z ) }.  
\end{eqnarray}
${\bf k}$ and $ \mu$ are the electron momentum and its  chemical potential, respectively.   $ \varepsilon_\text{c}({\bf k}) =  E_\text{g} + {\hbar^2 k^2}/{2 m_c} $ is the energy dispersion of the electron in  the conduction band and  $ \varepsilon_\text{v}({\bf k}) =   {\hbar^2 k^2}/{2  m_\text{v}} $  is the corresponding one  in    the valence band.

We emphasize that it is important to include the dynamics  in the Coulomb interaction and not only in the Green's functions of the electron and hole. To illustrate this importance, we evaluate the self-energy     by using   a static potential $V({\bf q})$ instead of $V({\bf q}, z-z')$.  Equation~(\ref{eq:DynaSelfE}) in this case  becomes frequency-independent $\Sigma({\bf k}) = -\beta^{-1}\sum_{\bf q} V({\bf q})  \sum_{z'} G({\bf k+ q}, z' )   \equiv  \sum_{\bf q} V({\bf q}) \left(  f({\bf k+ q} ) - \frac{1}{2} \right) $  where $f(k)$ is the Fermi-Dirac distribution function. In the zero-temperature limit and at charge  neutrality, the self-energy of the electron in the conduction  or valence bands is 
\begin{equation}
\Sigma_{\text{c/v}}({\bf k}) = \mp \frac{1}{2} \sum_{\bf q} V({\bf q}).
\label{Eq:NonDynaBGR}
\end{equation}
Namely, one  gets a rigid energy shift of the bands which is essentially  the BGR in the static limit. By neglecting dynamical effects in the potential, we lose the frequency dependence of the self-energies  and decouple the electron self-energy from that of the hole (i.e., the electron and hole have independent self-energies which are identical to the ones of free particles in the bands).   We will show later that the coupling and  dynamical self-energies are important to get the correct energy blueshift of the exciton resonance, as observed in experiment \cite{Amine_XYZ24}.

 One difficulty of self-energy calculations is the divergence of the sum over $\bf q$. We  illustrate this point by using the non-dynamical BGR in Eq. (\ref{Eq:NonDynaBGR}).  The 2D potential $V(q) $ scales as $q^{-2}$ in the short wavelength limit, resulting in a logarithmic divergence of the sum over $\bf q$.    We circumvent this problem by choosing a reference TMD structure with respect to which energy shifts are calculated.  Without loss of generality, we choose a reference system  whose corresponding potential  $V_0(q)$ is evaluated  with the following fixed dielectric constants  $\epsilon_\text{t}  = \epsilon_\text{b} = 3.8$.   The BGR of a given system with respect to the reference system is then given by 
\begin{eqnarray}
\tilde{\Sigma}_{\text{c/v}}({\bf k}) &=& \mp \frac{1}{2} \sum_{\bf q}  \left( V({\bf q}) - V_0({\bf q}) \right)  \nonumber \\
&=&\mp \frac{e^2}{2} \int_0^\infty dq \left( \frac{1}{\epsilon(q) } - \frac{1}{\epsilon^0(q) } \right). \label{Eq:BGRNonDyna}
\end{eqnarray}
$\epsilon(q)$ and $\epsilon^0(q)$ are the non-dynamical dielectric functions of the investigated TMD and  reference system, respectively (see Eq.~(\ref{Eq:DiFv2})). When dynamical effects are included, the self-energy in Eq.~(\ref{eq:DynaSelfE})  has the same divergence problem, and it can be regularized in a similar way
\begin{eqnarray}
\!\!\!\!\!\!\!\!\!\!\!\!\tilde{\Sigma}_i({\bf k},z ) &=& - \frac{1}{\beta} \sum_{{\bf q},z'} G_i({\bf k+ q}, z' ) \nonumber \\ &&\qquad \qquad \times \left[ V({\bf q}, z - z') -V_0({\bf q}) \right], \,\,\,\,\, \,\,\,\, \,\,\,\
\label{eq:DynaSelfE1}
\end{eqnarray} 
where the free electron Green's function now becomes
\begin{eqnarray}
G_i({\bf k},z )= \frac{1}{z - \varepsilon_i({\bf k}) + \mu- \tilde{\Sigma}_i({\bf k},z ) }.
\label{eq:GreenE} 
\end{eqnarray}
The dynamical self-energy $\tilde{\Sigma}_i({\bf k},z )$ can be self-consistently  calculated  from Eqs.~(\ref{eq:DynaSelfE1}) and (\ref{eq:GreenE})  using the iterative method (see Appendix \ref{app:discrete} for details). 

%%%%%%%%%%%%%%%%%%%%%%%%%%%%%%%%%%%%%%%%%%%%%%%%%%%%%%%%%%
%%%%%%%%%%%%%%%%%%%%%%%%%%%%%%%%%%%%%%%%%%%%%%%%%%%%%%%%%%
\begin{figure}[b]
\includegraphics[width=8.5cm]{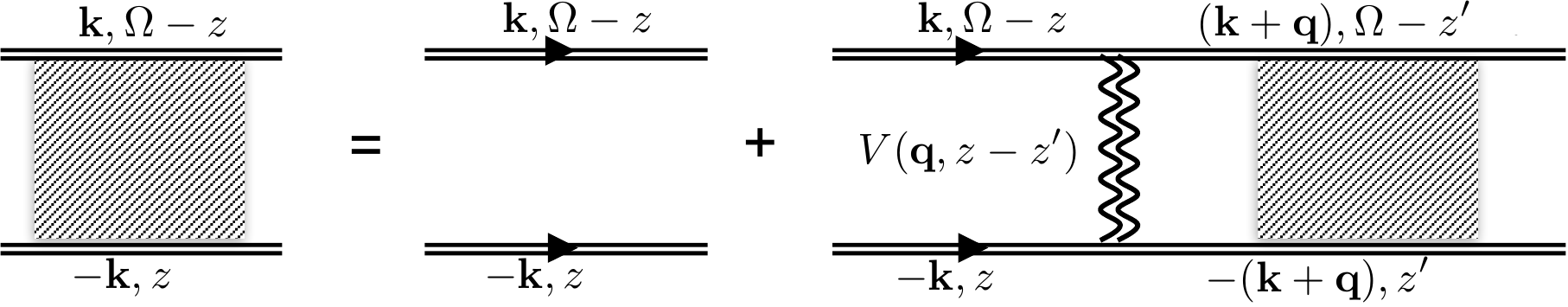}
\caption{ Feynman diagram for the Bethe-Salpeter Equation describing the interaction between electron (top double lines) and hole (bottom double lines) via the dynamical potential $V({\bf q}, z-z')$,  represented by the doubled wavy line. }\label{fig:BSE} 
\end{figure}

\subsection{ Dynamical Bethe-Salpeter Equation }

The BSE is an equation for bound states between two particles. Its  dynamical version is used here to describe the  interaction between electron and hole excited by light with negligible momentum. The  equation with its  Feynman diagram shown in Fig.~\ref{fig:BSE} reads \cite{Scharf_JPCM19,Haug_SchmittRink_PQE84,VanTuan_PRX17}
  
%\begin{widetext}
\begin{eqnarray}  \label{Bethe-Salpeter1}
 && \!\!\!\! \!\!\!\! \!\!\!\!G({\bf k},z,\Omega)= G^0({\bf k},z,\Omega) \,\,\, + \\ && \,\,\,\,\frac{1}{\beta}\sum_{{\bf  q},z'}G^0({\bf k},z,\Omega) V({\bf q},z-z')G({\bf k+q},z',\Omega) \nonumber, 
\end{eqnarray}
%\end{widetext}
where the Green's function of a free electron-hole pair is given by 
\begin{eqnarray}
 G^0({\bf k},z,\Omega) &=&  \frac{1}{\Omega - z - \varepsilon_\text{c}({\bf k})  - \tilde{\Sigma}_\text{c}({\bf k},\Omega - z )   + \mu }  \nonumber \\ &&\,\times  \frac{1}{ z + \varepsilon_\text{v}({\bf k})+  \tilde{\Sigma}_\text{v}({\bf k}, z ) - \mu} . 
 \label{Eq:FreeGreen}
\end{eqnarray} 
$\Omega$,  an even (bosonic) imaginary Matsubara energy, is related to the energy of the photon exciting the electron-hole pair. Equation~(\ref{Bethe-Salpeter1}) can be solved using the iterative method, the same method as the one used for calculating the self-energies in   Eqs.~(\ref{eq:DynaSelfE1})-(\ref{eq:GreenE}).  The details  are presented in Appendix \ref{app:discrete}.  One can notice that solutions of different bosonic frequencies $\Omega$ are decoupled, and therefore, equations of different $\Omega$s can be solved independently. The solutions are  then used to find the contracted pair function 
\begin{equation}
g({\bf k},\Omega) = -\beta^{-1} \sum_{z} G({\bf k},z,\Omega). 
\label{Eq:BSECon}
\end{equation}
The final step is to analytically continue the contracted pair function to the real-frequency axis, $g({\bf k},\Omega \rightarrow \omega + i \delta)$, using the Pad\'{e} approximation technique  \cite{VanTuan_PRX17,Scharf_JPCM19,Vidberg_JLTP1977}, as discussed in Appendix \ref{app:Pade}.     The real-frequency pair function is related to optical absorption by 
\begin{equation}
\text{A}(\omega) = - \sum_{{\bf k}} \text{Im} \left[ g({\bf k},\Omega \rightarrow \omega + i \delta)\right],
\label{Eq:Ana1}
\end{equation}
where $\delta$ is broadening parameter which might include effects  of finite exciton lifetime, scattering off  impurities, and thermal fluctuations. 
Note that temperature in this formalism sets the energy resolution of Matsubara frequencies and is not related to the broadening of resonance peaks which is controlled by $\delta$. In this work, we keep $\delta = 3$~meV for the sake of simplicity. More details are mentioned in Appendix \ref{App:NumAsp}.

In the non-dynamical regime, the potential and  self-energies are frequency independent.  The BSE in Eq.~(\ref{Bethe-Salpeter1}) can be further contracted, yielding 
\begin{equation}
 \!\!\! g({\bf k},\Omega)= g^0({\bf k},\Omega)- \sum_{{\bf  q}}g^0({\bf k},\Omega)\,\,  V({\bf q})\,\, g({\bf k+q},\Omega), 
 \label{Eq:StaBSE}
\end{equation} 
and the corresponding function of a free electron-hole pair is given by \cite{MahanBook,Haug_SchmittRink_PQE84}
\begin{eqnarray}
g^0({\bf k},\Omega) &=& -\beta^{-1} \sum_{z} G^0({\bf k},z,\Omega) \nonumber \\
&=& \frac{ f_{\text{v}}({\bf k}) - f_{\text{c}}({\bf k})}{\Omega    + \varepsilon_\text{v}({\bf k}) +  \tilde{\Sigma}_\text{v}({\bf k}) - \varepsilon_\text{c}({\bf k}) - \tilde{\Sigma}_\text{c}({\bf k})  }. \,\,\,\,\,\,\,\,\,\, \,\,\,\, \label{Eq:Free}
\end{eqnarray}
$f_{\text{c(v)}}({\bf k})$ is the Fermi-Dirac distribution function for electrons in the conduction (valence) band. Note that the free-pair function has the same form as the polarization function in the random-phase approximation. The only difference is that we have interband electron-hole excitation here instead of intraband particle-hole excitation in the polarization function. The analogy means that the method developed in this work is applicable for studying the dynamical response in the dielectric screening formalism.   

%%%%%%%%%%%%%%%%%%%%%%%%%%%%%%%%%%%%%%%%%%%%%%%
%%%%%%%%%%%%%%%%%%%%%%%%%%%%%%%%%%%%%%%%%%%%%%%
%%%%%%%%%%%%%%%%%%%%%%%%%%%%%%%%%%%%%%%%%%%%%%%
%%%%%%%%%%%%%%%%%%%%%%%%%%%%%%%%%%%%%%%%%%%%%%%
%%%%%%%%%%%%%%%%%%%%%%%%%%%%%%%%%%%%%%%%%%%%%%%
%%%%%%%%%%%%%%%%%%%%%%%%%%%%%%%%%%%%%%%%%%%%%%%
\section{RESULTS AND DISCUSSIONS} \label{sec:results}

We  first show  signatures of  dynamical effects in the BGR and in self-energies of free particles.  The self-energies are then used in  calculations of the pair function $g({\bf k},\Omega)$, which in turn is used to extract the spectral position of the charge-neutral exciton resonance. %The results point to the fact that while  self-energies of the exciton's electron and hole in  TMD monolayers supported on ionic crystals are mainly controlled by $\epsilon_\infty$,  the dominant contribution to the exciton binding energy and  BGR of free particles are mainly controlled by $\epsilon_0$.  

%%%%%%%%%%%%%%%%%%%%%%%%%%%%%%%%%%%%%%%%%%%%%%%
%%%%%%%%%%%%%%%%%%%%%%%%%%%%%%%%%%%%%%%%%%%%%%%
\subsection{Dynamical effects in the self-energy and BGR} \label{Result:BGR}

\begin{figure}[t] 
\centering
\includegraphics[width=8.5cm]{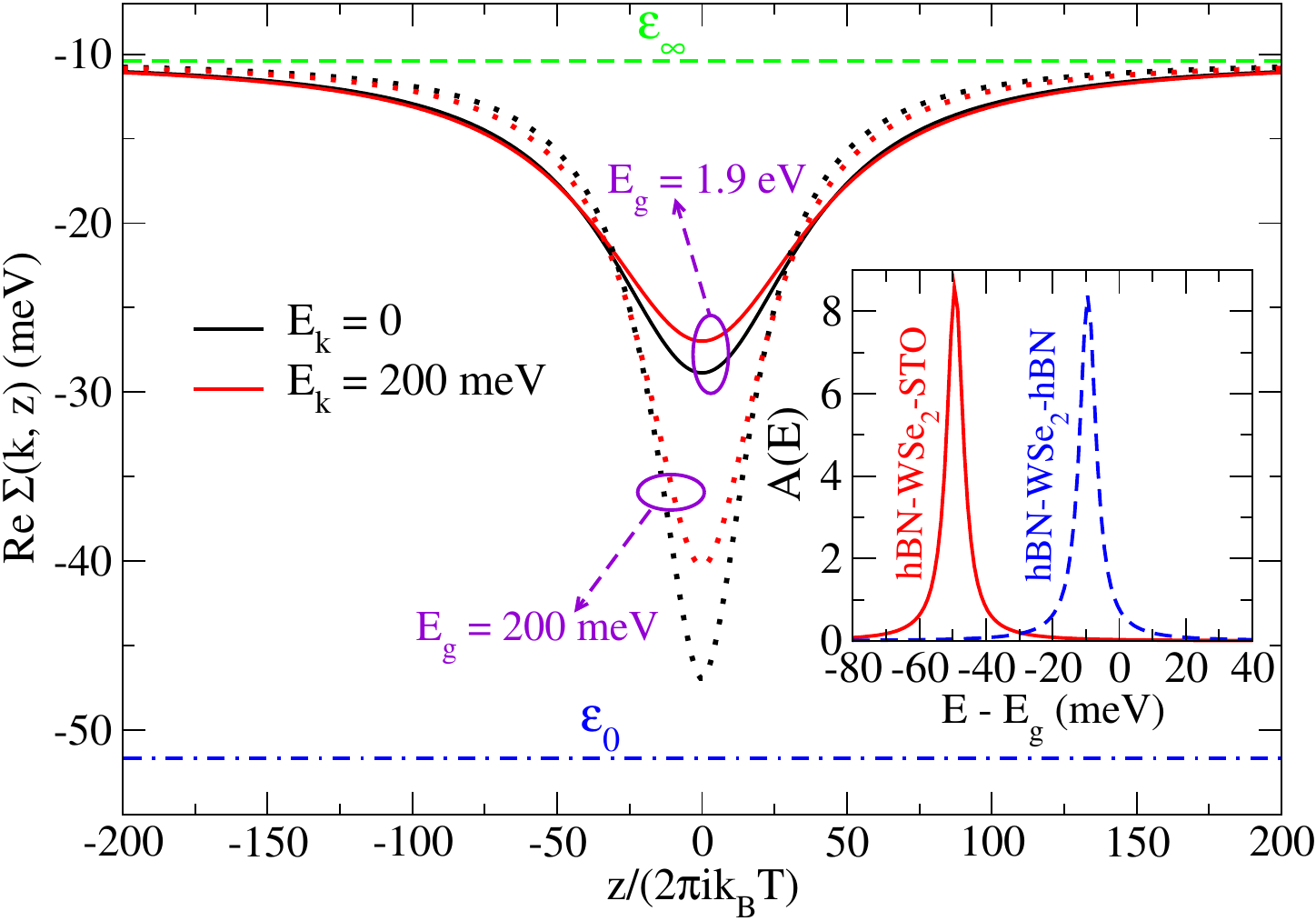}
\caption{ Real part of the self-energy $\Sigma_\text{c}({\bf k},z)$  of conduction-band electrons in hBN-WSe$_2$-STO as a function of Matsubara frequency $z= (2\ell+1)\pi i k_\text{B}T$. Solid (dotted) lines are self-consistent solutions of Eqs.~(\ref{eq:DynaSelfE1})-(\ref{eq:GreenE}) when $E_g=1.9$~eV (0.2~eV). For comparison, the dashed (dash-dotted) line is the solution of Eq.~(\ref{Eq:BGRNonDyna}) with $\epsilon_\infty$ ($\epsilon_0$). Inset: Spectral function of an electron at the edge of the conduction band, $E_k=0$, in hBN-WSe$_2$-STO (red solid line) and  hBN-WSe$_2$-hBN (blue dashed line). The resonance energy of the latter (former) emerges at $-9.5$~meV  ($-49.5$~meV), denoting the BGR with respect to the  edge of the conduction band in the non-dynamical reference system ($\epsilon_\text{b}= \epsilon_\text{t} = 3.8$).}\label{fig:BGR} 
\end{figure}

Figure \ref{fig:BGR} shows the self-energy of electrons $\Sigma_\text{c}({\bf k},z)$ as a function of Matsubara frequency $z$ for two representative energies, $E_{\bf k} = {\hbar^2k^2}/{2m_\text{c}} = \{0, 200 \text{ meV} \}$.    Dynamical effects are included by using the potential $V(q,z-z')$ in Eq. (\ref{Eq:DynaPoten}) adjusted to describe WSe$_2$ monolayer between STO and hBN layers. The parameters used in the calculation are listed in Appendix \ref{app:Para}. The bandgap energies in these calculations are $E_\text{g} =1.9$~eV (solid lines) and $E_\text{g} =0.2$~eV (dotted lines), where the chemical potential is set at the midgap, i.e., $\mu = E_\text{g}/2$. The  corresponding self-energies in the valence band, $\Sigma_\text{v}({\bf k},z)$, are almost the same as $\Sigma_\text{c}({\bf k},z)$ but with opposite sign (not shown).   This similarity  is reasoned  by similar  effective masses of electrons and holes, which together with the midgap chemical potential, result in a good electron-hole symmetry.  The self-energy is used to obtain the electron Green's function  $G_\text{c}({\bf k},z )$ through Eq.(\ref{eq:GreenE}). The conduction-band BGR is evaluated from the resonance energy of the spectral function  
\begin{equation}
A(E) = - \text{Im} \left[G_\text{c}({\bf k}=0,z \rightarrow E+ i\delta ) \right].
\label{Eq:Ana2}
\end{equation}
The analytical continuation from imaginary to  real energies $( z \rightarrow E)$ is performed by using the Pad\'{e} approximation technique (see Appendix \ref{app:Pade}). 

The inset of Fig. \ref{fig:BGR} shows the spectral functions $A(E)$ at the edge of the conduction band, $k=0$, in hBN-WSe$_2$-STO  and  hBN-WSe$_2$-hBN structures when   dynamical effects are included. The resonance emerges at the conduction-band edge, where the zero energy level corresponds to the edge of the conduction band in the non-dynamical reference system ($\epsilon_\text{t}  = \epsilon_\text{b} = 3.8$).  Table \ref{tab:BGREps} summarizes the BGR calculations using low- and high-frequency dielectric constants as well as  the dynamical dielectric function. The dynamical result is very close to the one calculated with $\epsilon_0$, providing strong evidence  that  charge particles in the monolayer are also screened by phonons.   On the other hand, the self-energies $\Sigma_\text{c}({\bf k},z)$ shown by solid lines in Fig. \ref{fig:BGR}  are closer to the self-energy calculated by  Eq.~(\ref{Eq:NonDynaBGR}) with $\epsilon_\infty$  (dashed line).  This resemblance  comes from the inclusion of the bandgap energy $E_\text{g}$. This inclusion does not affect the BGR because after analytical continuation $z \rightarrow E_\text{g} + E$ in Eq.~(\ref{Eq:Ana2}),   the bandgap energy in the denominator of Eq.~(\ref{eq:GreenE}) is eliminated (i.e., it is merely a reference energy level).

\begin{table}[t]
\caption{\label{tab:BGREps} BGR calculations with different dielectric parameters. The energies shifts are with respect to the reference system (hBN-WSe$_2$-hBN with $\epsilon(\omega)=\epsilon_\infty$ in the hBN layers).}

\begin{center}
\begin{tabular}{||c || c | c ||} 
 \hline
 \backslashbox{$\epsilon$}{Materials} & hBN-WSe$_2$-STO  & hBN-WSe$_2$-hBN  \\ [0.5ex] 
 \hline\hline
$\epsilon(\omega)$   & $-49.5$ meV  & $-9.5$  meV \\ 
 \hline
$\epsilon_0$ & $-51.6$ meV  & $-10.9$  meV  \\ 
 \hline
$\epsilon_\infty$  & $-10.4$ meV  & $0$  meV \\ 
 \hline

 \hline
\end{tabular}
\end{center}

\end{table}

\begin{figure*} 
\centering
\includegraphics[width=14cm]{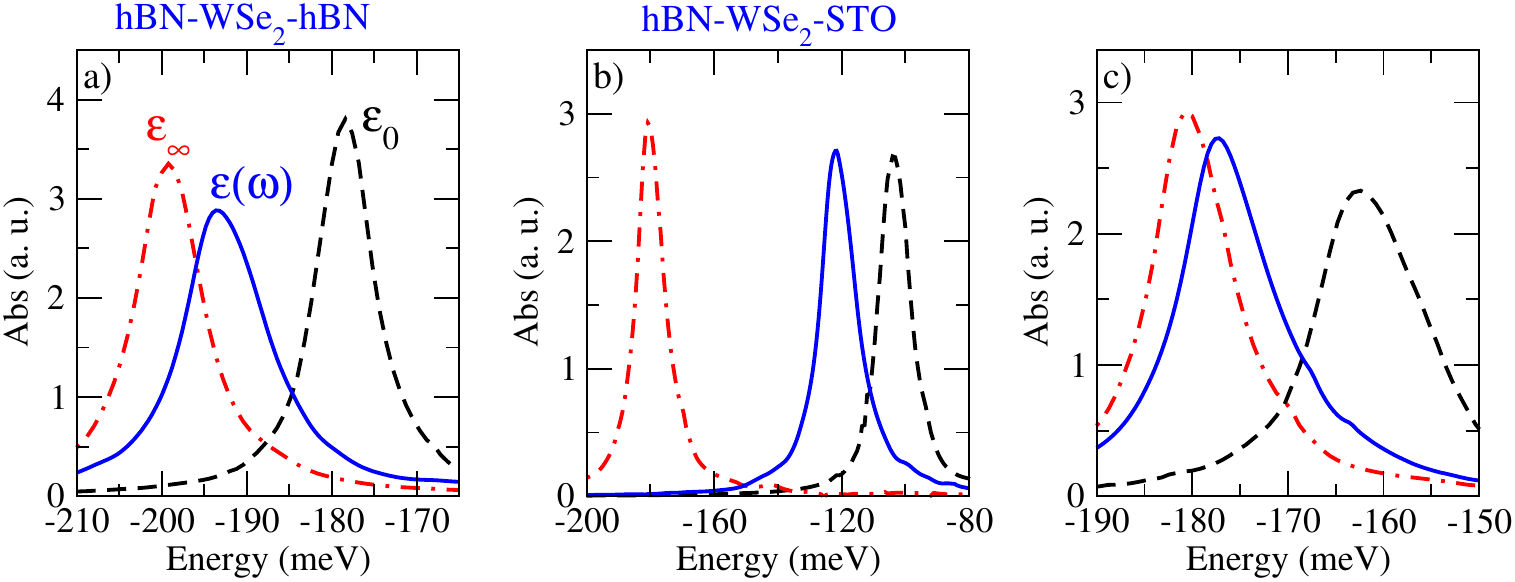}
\caption{  Absorption spectra of (a)  hBN-WSe$_2$-hBN and  (b) hBN-WSe$_2$-STO structures. Blue solid lines are results calculated with the dynamical potential $V({\bf q},\omega)$ but without self-energies of the electron and hole in the exciton. Black dashed lines (red dash-dotted lines) are respective results with the non-dynamical potential $V({\bf q})$ when using $\epsilon_0$ ($\epsilon_\infty$). (c)  Results of the same calculations as in (b) but when modeling the STO layer with $r_2 = 1.3$ instead of 4096.}\label{fig:AbsNoSelfE} 
\end{figure*}

%%%%%%%%%%%%%%%%%%%%%%%%%%%%%%%%%%%%%%%%%%%%%%%
%%%%%%%%%%%%%%%%%%%%%%%%%%%%%%%%%%%%%%%%%%%%%%%
\subsection{Dynamical effects in the optical spectrum}

%In the next section we  survey  effects of dynamical self-energies on the spectral position of neutral exciton which includes both electron and hole components. In contrast to what obtained above for BGRs of free charge particles, the involvement of both electron and hole self-energies  in exciton makes either electron or hole or both self-energies close to the high-frequency BGR. 

The dynamical potential  $V({\bf q}, z-z')$ affects both the BGR and binding energy. To decouple these effects, we focus first on the binding energy by neglecting the self-energy terms in the BSE (i.e., $\tilde{\Sigma}_\text{c/v}({\bf k},z ) = 0$ in Eq.~(\ref{Bethe-Salpeter1})).  The resulting absorption spectra of hBN-WSe$_2$-hBN and   hBN-WSe$_2$-STO structures are shown in Figs.~\ref{fig:AbsNoSelfE}(a) and (b), respectively. For comparison, we have also calculated the absorption spectra with the non-dynamical potential $V({\bf q})$ using low-frequency dielectric constants (dashed lines)  and high-frequency ones (dash-dotted lines). The exciton binding energy is $E_\text{b} = 122 $~meV  when using the dynamical potential in hBN-WSe$_2$-STO. Corresponding values of the non-dynamical calculations are $E_\text{b}^0 =104$~meV and $E_\text{b}^\infty = 181$~meV.  The dynamical binding energy is closer to the one calculated with $\epsilon_0$,  meaning that screening of the interaction between the electron and hole  is dominated by the low frequency part of the dielectric function. On the other hand, Fig.~\ref{fig:AbsNoSelfE}(a) shows opposite trend for the hBN-WSe$_2$-hBN structure, where the dynamical calculation is closer to the non-dynamical one calculated with $\epsilon_\infty$.

 To understand this behavior, we define a low-frequency contribution factor
\begin{equation}
\alpha \equiv \frac{E^\infty_b - E_\text{b}}{E^\infty_b - E^0_\text{b}}, 
\end{equation}
where $\alpha = 1$   ($\alpha = 0$) means that the exciton binding is completely controlled by $\epsilon_0$ ($\epsilon_\infty$). Using the calculated results in  Fig.~\ref{fig:AbsNoSelfE}, the low-frequency contribution factor is $\alpha = 0.77$ for hBN-WSe$_2$-STO and $\alpha =0.27$ for hBN-WSe$_2$-hBN. The difference stems from the Lyddane–Sachs–Teller relation of hBN and STO, expressed through the parameter $r_2$  in Eq.~(\ref{Eq:r2}). A larger $r_2$ leads to stronger dynamical contribution in Eq.~(\ref{Eq:DynaDielec}), which in turn pushes the dielectric function away from $\epsilon_\infty$ and closer to  $\epsilon_0$. Indeed, $r_2 > 10^3$ in low-temperature STO versus $r_2 \approx 1.3$ in hBN. To confirm that our understanding is correct, we repeated the calculation  of hBN-WSe$_2$-STO but with changing $r_2$ of STO from $4096$ to $1.3$. This change corresponds to lowering  $\epsilon_0$ of STO from $25000$ to $7.8$.  The result of the calculation is shown in Fig.~\ref{fig:AbsNoSelfE} (c). We obtain  $\alpha = 0.2$, confirming that the contribution of $\epsilon_\infty$ is indeed dominant at smaller $r_2$. % The results show that  the dielectric variation ratio $(r_2-1) =  (\epsilon_0 - \epsilon_\infty)/\epsilon_\infty $ is the factor controlling the dynamical effects. This ratio is very large  in  STO or TiO$_2$, leading to energy blueshift of the exciton resonance compared with  hBN-WSe$_2$-hBN when we also consider the effect from dynamical self-energies, as we explain next.  

\begin{figure*}[t] 
\centering
\includegraphics[width=14cm]{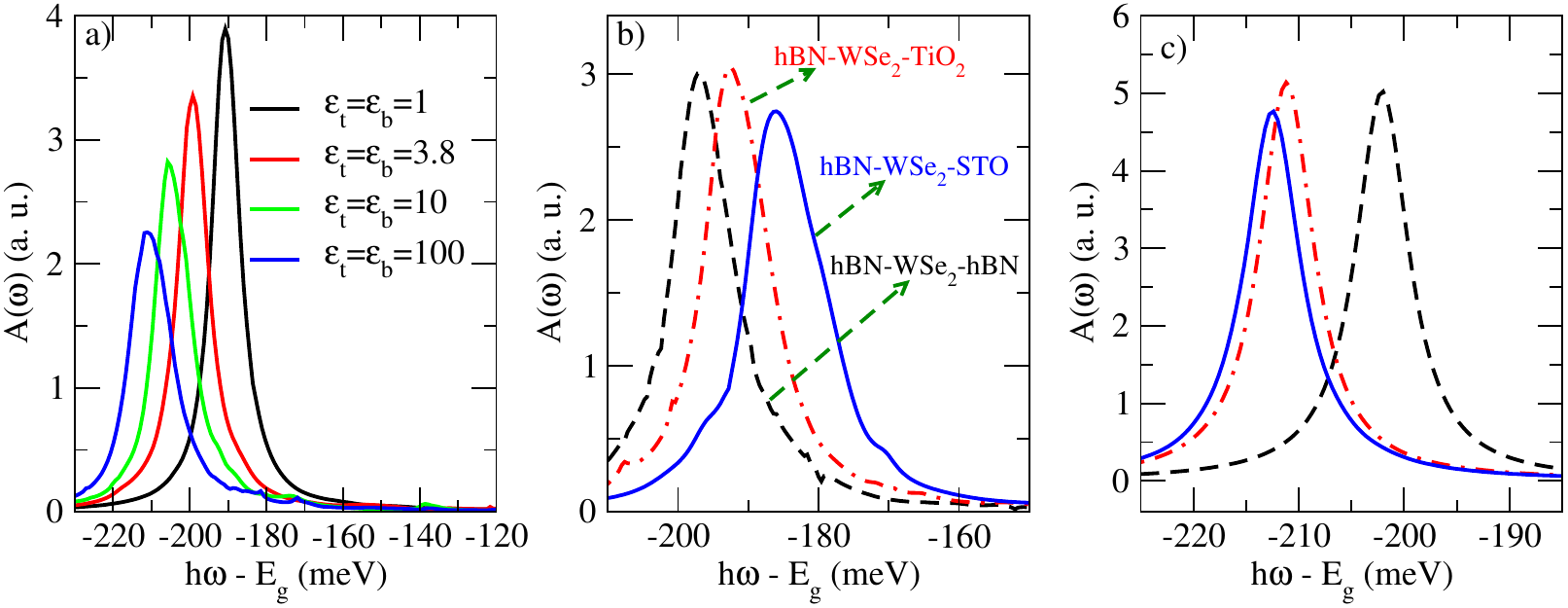}
\caption{ (a) Calculated absorption spectra with the non-dynamical potential of various dielectric environments. The BGR effect is calculated with respect to the reference system ($\epsilon_b=\epsilon_t=3.8$). The exciton resonance shows energy redshift when the dielectric screening is increased. (b) Calculated absorption spectrum with the dynamical potential and $E_\text{g} = 1.9$~eV. The exciton resonance shows energy blueshift when the dielectric screening of the environment is increased.  (c) The same as in (b) but for $E_\text{g} = 0.2$~eV, showing an energy redshift trend. }\label{fig:DynaAbs} 
\end{figure*}

%
%The low- and high-frequency regimes for a material is set by the transverse optical phonon frequency $\hbar \omega_{\text{TO}}$. If the average transferred energy is much lower than  $\hbar \omega_{\text{TO}}$ then it is likely that the binding energy in the material is controlled by LFD function.  hBN is a material with larger transverse optical phonon $\hbar \omega_{\text{TO}} $ compared to the corresponding one in STO (92 meV \cite{Cai_SSC2007} versus 16 meV \cite{Evarestov_PRB2011}). This means that   the binding energy  in hBN-encapsulated sample is more controlled by lower-frequency dielectric function than one in  hBN-WSe$_2$-STO. 

%The band-gap energy of the monolayer gets smaller when the effective dielectric screening is stronger due to weakening of the Coulomb potential at the origin ($r \rightarrow 0$), namely at the position of a charge particle \cite{Haug_SchmittRink_PQE84,Marauhn_PRB23}. On the other hand, the electron-hole binding energy is affected by the change to the Coulomb potential at finite distances and not only at $r=0$. Since the difference between the Coulomb potentials of  two dielectric environments is largest at $r=0$, the reduction in band-gap energy cannot be completely offset by weaker binding energy, leading to a small overall redshift of the resonance energy when the dielectric screening increases, $\Delta\,E_{X_0} = \Delta E_g -  \Delta E_b \lesssim 0$, as shown in Eq. (\ref{eq:red}).  

Next, we include the self-energy in calculations of the absorption spectra and consider the competition  between BGR and binding energy.   The  exciton resonance energy is the difference between band-gap energy and electron-hole binding energy, $E_{X_0} = E_g - E_b$. When the dielectric environment surrounding the monolayer or quantum well changes, the renormalized bandgap and binding energies tend to offset each other, such that $E_{X_0}$ is only slightly changed \cite{Haug_SchmittRink_PQE84,Raja_NatComm17,Scharf_JPCM19}. For example, a stronger dielectric screening leads to smaller band-gap energy and smaller binding energy, where their difference is much smaller than the change of each of them.  As mentioned in Sec.~\ref{sec:Mov}, a common observation in monolayer semiconductors is a small overall energy redshift of $E_{X_0}$ when the materials that encapsulate the monolayer are replaced with higher-dielectric constant materials. Figure \ref{fig:DynaAbs}(a) shows the calculated absorption spectra from the BSE using a non-dynamical potential. The  BGR effect is calculated with respect to the reference system ($\epsilon_b=\epsilon_t=3.8$).  In accordance with the analysis of Sec.~\ref{sec:Mov}, the exciton resonance redshifts when  the dielectric constant of the surrounding environment increases.

Yet, the energy redshift can turn to a blueshift if dynamical effects are considered through the potential $V({\bf q}, z-z')$ and self-energies $\tilde{\Sigma}_\text{c/v}({\bf k},z )$, and if replacing the encapsulating materials involve a large change in $\epsilon_0$ and a small change in $\epsilon_\infty$. Figure~\ref{fig:DynaAbs}(b) shows the resulting absorption spectra of three different structures:  hBN-WSe$_2$-hBN,  hBN-WSe$_2$-TiO$_2$, and  hBN-WSe$_2$-STO.  In agreement with experimental results \cite{Amine_XYZ24}, replacing the supporting hBN layer with TiO$_2$ leads to energy blueshift, which is further increased when STO is used as support. The opposite energy-shift trends of the non-dynamical and dynamical calculations in Figs.~\ref{fig:DynaAbs}(a) and (b), can be explained as follows. The binding energy is mainly dominated by the low-frequency part of the dielectric function, where the change is from $\epsilon^\text{hBN}_0 = 4.9$ to $ \epsilon^{\text{TiO}_2}_0 \sim 200$ and then to $\epsilon^\text{STO}_0= 25000$. As a result, the change in binding energy is relatively significant. On the other hand, the self-energies of the electron and hole in the exciton have larger contribution from the high-frequency part, where the change is from $\epsilon^\text{hBN}_\infty = 3.8$ to $\epsilon^\text{TiO$_2$}_\infty \sim \epsilon^\text{STO}_\infty \sim 6$. As a result, the BGR effect is relatively mitigated. The confluence of both trends is that the energy redshift from BGR is smaller than the energy blueshift  from binding energy ($  \Delta  E_\text{g} < |\Delta E_\text{b}|$), leading to overall energy blueshift of the exciton resonance.  

We provide mathematical and physical reasonings to the observed energy blueshift (i.e., why weakening of the binding energy is stronger than the bandgap-energy shrinkage). As shown by Eq.~(\ref{Eq:FreeGreen}), the self-energy of the electron in the exciton is associated with $(\Omega - z)$ whereas that of the hole with $z$. The bosonic frequency $\Omega$ is related to the photon energy which is of the order of the bandgap energy; a large value compared with phonon or binding energies. Consequently, the self-energy of at least one of the exciton's components asymptotically approaches the value of the self-energy when calculated non-dynamically with $\epsilon_\infty$. An alternative view is that the energy difference between the exciton components is encoded as a  time-dependent phase factor $\exp(iE_g t/\hbar)$, which leads to a dominant contribution from the high-frequency part of the dielectric function to the exciton's BGR.

Finally, we recall that the edge of an energy band is merely a reference level when  the interest is in the BGR of a charge particle in this band. The energy reference level is eliminated in the analytical continuation step (see discussion at the end of Sec.~\ref{Result:BGR}). Therefore, the dominant contribution to the BGR of a charge particle comes from the low-frequency part of the dielectric function (Table~\ref{tab:BGREps}). On the other hand, the  reference energy level for a bound pair (exciton) mandates that at least its electron or hole are subjected to high frequencies that cannot be eliminated by analytical continuation. To confirm the importance of the bandgap energy, we have performed additional calculations with similar parameters to the ones used in Fig.~\ref{fig:DynaAbs}(b), but with a much smaller bandgap energy, $E_\text{g} = 0.2$~eV.  The results are shown in \ref{fig:DynaAbs}(c). As expected from the explanation above, the trend is now reversed and we observe a small energy redshift coming from lesser contribution of high frequencies to the BGR (Fig.~\ref{fig:BGR}).

%%%%%%%%%%%%%%%%%%%%%%%%%%%%%%%%%%%%%%%%%%%%%%%
%%%%%%%%%%%%%%%%%%%%%%%%%%%%%%%%%%%%%%%%%%%%%%%
%%%%%%%%%%%%%%%%%%%%%%%%%%%%%%%%%%%%%%%%%%%%%%%
%%%%%%%%%%%%%%%%%%%%%%%%%%%%%%%%%%%%%%%%%%%%%%%
\section{Summary} \label{sec:con}

%The model should also be somewhat applicable at low charge densities and higher temperatures, where the particles are itinerant but their wavefunctions overlap is negligible. The DOS function in this case can be extracted by averaging over many random distributions can recover the itinerant nature of the particles.   However, t

We have presented a model that incorporates dynamical dielectric screening effects in the Bethe-Salpeter Equation. The model allows one to study excitons in transition-metal dichalcogenide monolayers that are embedded in various dielectric environments. We have employed an iterative numerical technique to solve the Bethe-Salpeter Equation, allowing us  to perform comprehensive calculations with a large number of Matsubara frequencies and fine mesh in momentum space. The theory sheds light on the intricate energy shifts of exciton resonances. Assuming that the bandgap energy is evidently larger than the exciton binding energy, we can have one of two opposite trends when the materials on top and/or below the monolayer are replaced  with higher-dielectric constant materials. If the involved materials are such that their low- and high-frequency dielectric constants are not evidently different ($\epsilon_0/\epsilon_\infty \not\gg 1$), then we should expect the exciton energy to redshift in the new environment. On the other hand, the exciton energy should blueshift if $\epsilon_0/\epsilon_\infty \gg 1$. These findings, identified by the inclusion of dynamical dielectric screening, help us to explain recent measurements in hBN-TMD-hBN, hBN-TMD-TiO$_2$, and hBN-TMD-STO devices \cite{Amine_XYZ24}. 

Beyond the agreement of the theory with experiment, the analysis identifies a distinction between the bandgap energy renormalization of excitonic complexes and the energy renormalization of free electrons in the conduction band or free holes in the valence band.   In the excitonic case, the renormalization has important contribution from the high-frequency part of the dielectric function. On the other hand, energy renormalizations of  free electrons in the conduction band and/or free holes in the valence band are dominated by the low-frequency part of the dielectric function. One important consequence is that if $\epsilon_0$ and  $\epsilon_\infty$ are much different, the bandgap energy renormalization that one measures in ARPES experiments cannot be used to infer the bandgap energy renormalization of bound electron-hole pairs. 

%\subsection*{Outlook} 

%https://www.nature.com/articles/s41586-020-2085-3

\acknowledgments{ This work is supported by the Department of Energy, Basic Energy Sciences, Division of Materials Sciences and Engineering under Award No. DE-SC0014349.}

%%%%%%%%%%%%%%%%%%%%%%%%%%%%%%%%%%%%%%%%%%%%%%%%%%%%%%%%%%%%%%%%%%%%%%%%%%%%%%%
%%%%%%%%%%%%%%%%%%%%%%%%%%%%%%%%%%%%%%%%%%%%%%%%%%%%%%%%%%%%%%%%%%%%%%%%%%%%%%%
%%%%%%%%%%%%%%%%%%%%%%%%%%%%%%%%%%%%%%%%%%%%%%%%%%%%%%%%%%%%%%%%%%%%%%%%%%%%%%%
%%%%%%%%%%%%%%%%%%%%%%%%%%%%%%%%%%%%%%%%%%%%%%%%%%%%%%%%%%%%%%%%%%%%%%%%%%%%%%%

\appendix
%%%%%%%%%%%%%%%%%%%%%%%%%%%%%%%%%%%%%%%%%%%%%%%%%%%%%%%%%%%%%%%%%%%%%%%%%%%%%%%
%%%%%%%%%%%%%%%%%%%%%%%%%%%%%%%%%%%%%%%%%%%%%%%%%%%%%%%%%%%%%%%%%%%%%%%%%%%%%%%
\section{Coulomb potential} \label{ap:Ponten}

We simulate the structure geometry as a TMD monolayer with thickness $d$ sandwiched between top and bottom layers with dielectric constants $\epsilon_\text{t}$ and $\epsilon_\text{b}$.  
The TMD monolayer is modeled as three atomic sheets with polarizabilities $ \chi_+$ for the central Tungsten (W) sheet  and $ \chi_-$  for the top and bottom Selenium (Se)  ones, displaced by $\pm d/4$ from the center. The model was developed in Ref. \cite{VanTuan_PRB18} and has been employed to study several problems \cite{VanTuan_PRL22,VanTuan_PRB22,VanTuan_PRL19}. The resulting static potential for the interaction between two charges in the monolayer is
\begin{equation}
V(q) = \frac{2 \pi e^2}{A \,\,  \epsilon(q) \,\, q},
\label{Eq:Static}
\end{equation}
where the  dielectric function reads
\begin{equation}\label{Eq:DiFv2}
\epsilon(q)=\frac{1}{2}\left[\frac{N_t(q)}{D_t(q)}+\frac{N_b(q)}{D_b(q)}\right].
\end{equation}
Defining $p_j \equiv (\epsilon_j-1)/(\epsilon_j+1)$ for the top and bottom dielectric constants ($j=b/t$), we get that 
\begin{eqnarray}\label{Eq:DiFv2def}
D_j(q) &=& 1+q\ell_- -q\ell_- (1+p_j)\text{e}^{-\frac{qd}{2}} - (1-q \ell_- ) p_j \text{e}^{-qd}, \nonumber \\
N_j(q) &=& \left(1+q\ell_-\right)\left(1+q\ell_+\right) \nonumber \\ & + & \left[\left(1-p_j\right)-\left(1+p_j\right)q\ell_+\right]q\ell_-\text{e}^{-\frac{qd}{2}} \nonumber \\
    &+& (1-q\ell_-)(1-q\ell_+ )p_j\text{e}^{-qd}.
\end{eqnarray}
where $\ell_\pm = 2 \pi  \chi_\pm$. 

The Coulomb potential in Eq.~(\ref{Eq:Static}) becomes frequency dependent by using dynamical polarization parameters,  $p_{j}(\omega) \equiv (\epsilon_{j}(\omega)-1)/(\epsilon_{j}(\omega)+1)$, where $\epsilon_{j}(\omega)$ is the dielectric function of the $j=b/t$ layer (Eq.~(\ref{Eq:DynaDielec})).

%%%%%%%%%%%%%%%%%%%%%%%%%%%%%%%%%%%%%%%%%%%%%%%%%%%%%%%%%%%%%%%%%%%%%%%%%%%%%%%
%%%%%%%%%%%%%%%%%%%%%%%%%%%%%%%%%%%%%%%%%%%%%%%%%%%%%%%%%%%%%%%%%%%%%%%%%%%%%%%
\section{Iterative method} \label{app:discrete}

The iterative method is used in calculations of Eqs.~(\ref{eq:DynaSelfE1})-(\ref{Bethe-Salpeter1}). These equations can generally be written as
 
 \begin{equation}
 \Gamma({\bf k},z ) =  \frac{1}{\beta} \sum_{{\bf q},z'}  F\left(z,\Gamma({\bf k+q},z') \right),
 \label{Eq:GenEq}
 \end{equation}
 where $\Gamma({\bf k},z )$ represents either  the self-energy $\tilde{\Sigma}_i({\bf k},z ) $ in Eqs.~(\ref{eq:DynaSelfE1})-(\ref{eq:GreenE})  or  the interacting pair Green's function   $G( {\bf k},z,\Omega)$ in Eq. (\ref{Bethe-Salpeter1}). We have omitted the bosonic frequency $\Omega$ in Eq.~(\ref{Eq:GenEq}) because the BSE of different $\Omega$ parameters are decoupled and can be solved separately.  When we are interested in self-energies,  $F\left(z, \Gamma({\bf k+ q},z') \right)$ is a composite function of the form  
 \begin{eqnarray}
 F\left(z, \Gamma({\bf k+q},z') \right) =    \frac{  V_0({\bf q})- V({\bf q}, z - z') }{{z - \varepsilon_i({\bf k + q}) + \mu- \Gamma({\bf k + q},z ) }}  \,\,\,  \,\,\, \,\,\, \,\,\,  \,\,\, 
 \end{eqnarray} 
%  \begin{equation}
% F\left(z, \Gamma({\bf k},z') \right) =    \frac{  V_0({\bf q})- V({\bf q}, z - z') }{{z - \varepsilon_i({\bf k }) + \mu- \Gamma({\bf k },z ) }.} 
% \end{equation}
 with $\Gamma({\bf k},z ) \equiv  \tilde{\Sigma}_i({\bf k},z ) $.  And when interested in the BSE

  \begin{eqnarray}
 F\left(z, \Gamma({\bf k+ q},z') \right) &=&  G^0({\bf k},z,\Omega) \,\, \left[\beta \delta_{\bf q,0} \delta_{z,z'}  \right. \nonumber \\
 &+& \left. V({\bf q}, z-z') \Gamma({\bf k+q},z',\Omega)  \right] \,\,\,  \,\,\, \,\,\, \,\,\, \,\,\,\,\,\, 
 \end{eqnarray}
 with  $\Gamma({\bf k},z ) \equiv G({\bf k},z,\Omega)  $.

The computational cost can be  greatly reduced  if one considers a two-dimensional system with circular symmetry in momentum space $\bf k$, i.e.,  $\Gamma({\bf k},z) \equiv \Gamma(k,z)$. Equation~(\ref{Eq:GenEq}) then becomes
%\begin{widetext}
\begin{eqnarray} 
\!\! \!\!\!\Gamma( k,z) = \sum_{z'}  \int_0^{\infty}  \frac{A \,\,  dq }{4 \pi^2}   \int_0^{\pi}   F\left(z,\Gamma({\bf k+q},z') \right)   d\theta.  \,  \,\,  \label{Eq:BSEFull1}
 \end{eqnarray}
% \end{widetext}
The equation is solved  numerically by  discretizing the momentum space, i.e., we  calculate   $\Gamma(k_i,z)$ at $n_K$ representative momenta $\{ k_i =\left( i - \frac{1}{2}\right)dk,  i = \overline{1,n_K}\}$ and $n_Z$ fermionic Matsubara frequencies.  Here, we divide the momentum space to $n_K$ rings with similar thickness $dk=K_\text{Max}/n_K$ where the cutoff momentum  $K_\text{Max} $  is chosen large enough to neglect  contributions from states above such momentum. The cutoff energy corresponding to $K_\text{Max}$ is $E^k_\text{Cut}$. The number of Matsubara frequencies $n_Z$ involved in the calculation guarantees that $E^z_\text{Cut} = n_Z\times  \pi k_\text{B} T $  is out of the accessible range of all related energy quantities of the considered phenomena (e.g.,  bandgap energy of the TMD monolayer, exciton binding energy, and kinetic energies of the electron and hole in the exciton).  

For  $({\bf k}_i+{\bf q})$ in Eq.~(\ref{Eq:BSEFull1}) to fall into the  $j^\text{th}$ ring, we set the condition
\begin{equation}
\mathcal{C}_{ i,j}: \,\,\, \left( j - 1\right) \le \frac{\sqrt{q^2 + k^2_{ i} + 2 k_{ i} q \cos \theta }}{dk}  < j.  
\end{equation}
Using this condition, we can rewrite Eq.~(\ref{Eq:BSEFull1}) in a form suitable for numerical calculation   
\begin{equation}  
 \Gamma(k_i,z) =  \sum_{z', j } \mathcal{F}_{i,j}(z,z')\,\,,\,\,\,\, 
 \label{BetheEq3}
 \end{equation}
where the function $ \mathcal{F}_{i,j}(z,z')$ is defined  by 
\begin{eqnarray} 
  \mathcal{F}_{i,j}(z,z')  = \frac{A }{4 \pi^2}    \int_{\{q,\theta\} \in  \mathcal{C}_{i,j}}     F\left(z,\Gamma({\bf k}_i+{\bf q},z') \right)      dq  d\theta.\,\,\,\,\,\,\,\, \,\,\,\, 
 \end{eqnarray}
This equation can be solved by matrix inversion \cite{VanTuan_PRX17,Scharf_JPCM19}. However, the computational cost of such calculation is expensive and prohibits the use of a large number of Matsubara frequencies and fine momentum mesh. Instead, we use an iterative method which helps to solve Eq.~(\ref{BetheEq3}) at a much smaller computational cost.  The iterative steps are
\begin{enumerate}
 \item Start  by guessing a solution, usually the non-interacting function $\Gamma_{\text{in}}(k_i,z) \equiv \Gamma^0(k_i,z)  $.
 \item  Substitute the input function $\Gamma_{\text{in}}(k_i,z)$ into the right hand side of Eq.~(\ref{BetheEq3})
 to find the output    function  $\Gamma_{\text{out}}(k_i,z)$ on the left hand side. 
 \item Use $\Gamma_{\text{out}}(k_i,z) $ as an input to the right-hand side of Eq.~(\ref{BetheEq3}) repeatedly until convergence is reached (i.e., $\Gamma_{\text{out}}(k_i,z) = \Gamma_{\text{in}}(k_i,z)  $).  
\end{enumerate}

In calculation of the interacting pair Green's function $G({\bf k},z,\Omega)$, the iterative procedure can be performed for  each Matsubara frequency $\Omega = \Omega_{j}$ independently.  The convergence of the iterative method can be sped up by using  the converged results of the higher frequency $G({\bf k},z,\Omega_{j})$ as the initial trial function for the next lower frequency $G_\text{in}({\bf k},z,\Omega_{j-1}) \equiv G({\bf k},z,\Omega_{j})$.

The converged solution  of the BSE equation   is difficult to obtain  for low values of $\Omega_j$. Fortunately, we can get rid of the the first few lowest  $\Omega_j$ frequencies because their Green's functions only contribute to states close to the continuum (e.g., 2$s$, 3$s$, and so on). The 1$s$ exciton state, with binding energy of hundreds of meV, is mostly controlled by high values of  $\Omega_j$. The convergence of the analytical continuation depends on the number of bosonic frequencies, $n_\Omega$,  as discussed in  Appendix \ref{App:NumAsp}.  In addition, convergence of the iterative method for the BSE can be improved by lowering the change from the input to output Green's functions. This can be done by modifying the third step above, where instead of directly using the output Green's function, $G_\text{out}({\bf k},z,\Omega_{j})$,   we use a linear combination of the old input and output Green's functions as input for the next iterative step
\begin{eqnarray}
\eta G_\text{in}({\bf k},z,\Omega_{j}) + (1- \eta) G_\text{out}({\bf k},z,\Omega_{j}) \rightarrow G_\text{in}({\bf k},z,\Omega_{j}). \,\,\,\, \,\,\,\,\, \,\,\,\,\,
\end{eqnarray} 
The calculations in this work use $\eta = 0.5$. 

One useful property  is that the contracted exciton Green's function obeys
\begin{equation}
g({\bf k},-\Omega) = g^*({\bf k},\Omega),
\end{equation}
thereby saving half the computational effort  by calculating $g({\bf k},\Omega)$ for positive Matsubara frequencies and using the above relation to find   $g({\bf k},\Omega)$ of negative Matsubara frequencies. 
 
Once the self-energy $\tilde{\Sigma}_i({\bf k},z ) $ and the contracted exciton pair function $g({\bf k},\Omega) $ are obtained, the Pad\'{e} approximation technique is employed to extract the BGR and absorption spectrum, respectively.  

%%%%%%%%%%%%%%%%%%%%%%%%%%%%%%%%%%%%%%%%%%%%%%%%%%%%%%%%%%%%%%%%%%%%%%%%%%%%%%%
%%%%%%%%%%%%%%%%%%%%%%%%%%%%%%%%%%%%%%%%%%%%%%%%%%%%%%%%%%%%%%%%%%%%%%%%%%%%%%%
\section{Pad\'{e} approximation technique} \label{app:Pade}

Pad\'{e} approximation is a technique used to perform  analytical continuation from imaginary Matsubara energies to real ones \cite{VanTuan_PRX17,Scharf_JPCM19,Vidberg_JLTP1977}. Usually the method is used for finding the real frequency Green's function when its values $\{G_i\}$ are known at $N$ Matsubara frequencies $\{ z_i\}$ (or $\{\Omega_i \}$ in case of bosonic Green's functions), where $i = \overline{1, N}$. In this work, the Green's function is obtained from Eqs.~(\ref{eq:GreenE}) or (\ref{Eq:BSECon}). To perform the continuation, we look for the Green's function of each momentum $\bf k$ in form of a rational fraction 
\begin{equation}
G_N(z)=\cfrac{a_1}{1+\cfrac{a_2(z-z_1)}{1+\cfrac{a_3(z-z_2)}{1+\cdots  \cfrac{\vdots}{ \cfrac{\vdots}{1+\cfrac{a_N(z-z_{N-1})}{1}\,\,,} }   }       }  }
\label{Pade1977}
\end{equation}
where the coefficients $a_i$ are to be determined so that 
\begin{equation}
 G_N(z_i)=G_i; \quad  i=\overline{1, N}.
 \label{Eq:Condi}
\end{equation}

If one defines a set of functions $\{f_j(z) \}$ for $ j=\overline{1, N}$ by the following  recursion 
\begin{eqnarray}
 f_1(z_i)&=&G_i; \quad \mbox{for} \quad  i=\overline{1, N}  \nonumber \\
 f_j(z) &= &\frac{f_{j-1}(z_{j-1})-f_{j-1}(z)}{(z-z_{j-1})f_{j-1}(z)},
 \label{Eq:Recur}
\end{eqnarray}
the coefficients $a_i$ in Eq. (\ref{Pade1977}) are given by 
\begin{equation}
 a_i=f_{i}(z_i) \quad \mbox{for}   \quad i=\overline{1, N}.
 \label{Eq:aiCoff}
\end{equation}
Indeed, the recursion in Eq .(\ref{Eq:Recur}) leads to 

\begin{equation}
 f_{j-1}(z)=\frac{f_{j-1}(z_{j-1})}{1+(z-z_{j-1})f_{j}(z)},
 \label{Eq:MRecur}
\end{equation}
which means that 
\begin{eqnarray}
 f_{1}(z)&=&\frac{f_{1}(z_{1})}{1+(z-z_{1})f_{2}(z)}.
\end{eqnarray}
The condition of $f_1(z_1) = G_1 = G_N(z_1) $ from Eqs.~(\ref{Eq:Condi})-(\ref{Eq:Recur}) leads to $a_1=f_{1}(z_1)$. Applying the recursion in Eq.~(\ref{Eq:MRecur}) one more time, we have
\begin{eqnarray}
 f_{1}(z)
 =\frac{a_1}{1+(z-z_{1}) \frac{f_{2}(z_{2})}{1+(z-z_{2})f_{3}(z)}  },  
\end{eqnarray}
which can be combined with the condition of $f_1(z_2) = G_2= G_N(z_2) $ from Eqs.~(\ref{Eq:Condi})-(\ref{Eq:Recur}) to prove that  $a_2=f_{2}(z_2)$. The same procedure can be performed with higher indices to prove Eq.~(\ref{Eq:aiCoff}).

The introduction of the recursive relation in Eq.~(\ref{Eq:Recur}) supports the calculation of the coefficients $\{a_i\}$ through the following steps

\begin{itemize}
  \item Start with  $f_1(z_i) = G_i$ for  $ i=\overline{1, N} $. 
  \item Express  $f_j(z_i)$ of higher indices $j>1$ by assigning $z=z_i$ in Eq.~(\ref{Eq:Recur}). %$ f_j(z_i) = \frac{f_{j-1}(z_{j-1})-f_{j-1}(z_i)}{(z_i-z_{j-1})f_{j-1}(z_i)},$ 
 \item Obtain $a_i$ from Eq.~(\ref{Eq:aiCoff}).  

\end{itemize}

The real-frequency Green's function is obtained from Eq.~(\ref{Pade1977}) by replacing $z$ with $\omega + i\delta$, where the broadening parameter $\delta$ takes into account effects of finite lifetimes, scattering off impurities, and thermal fluctuations. 

\begin{figure}[t] 
\centering
\includegraphics[width=8.5cm]{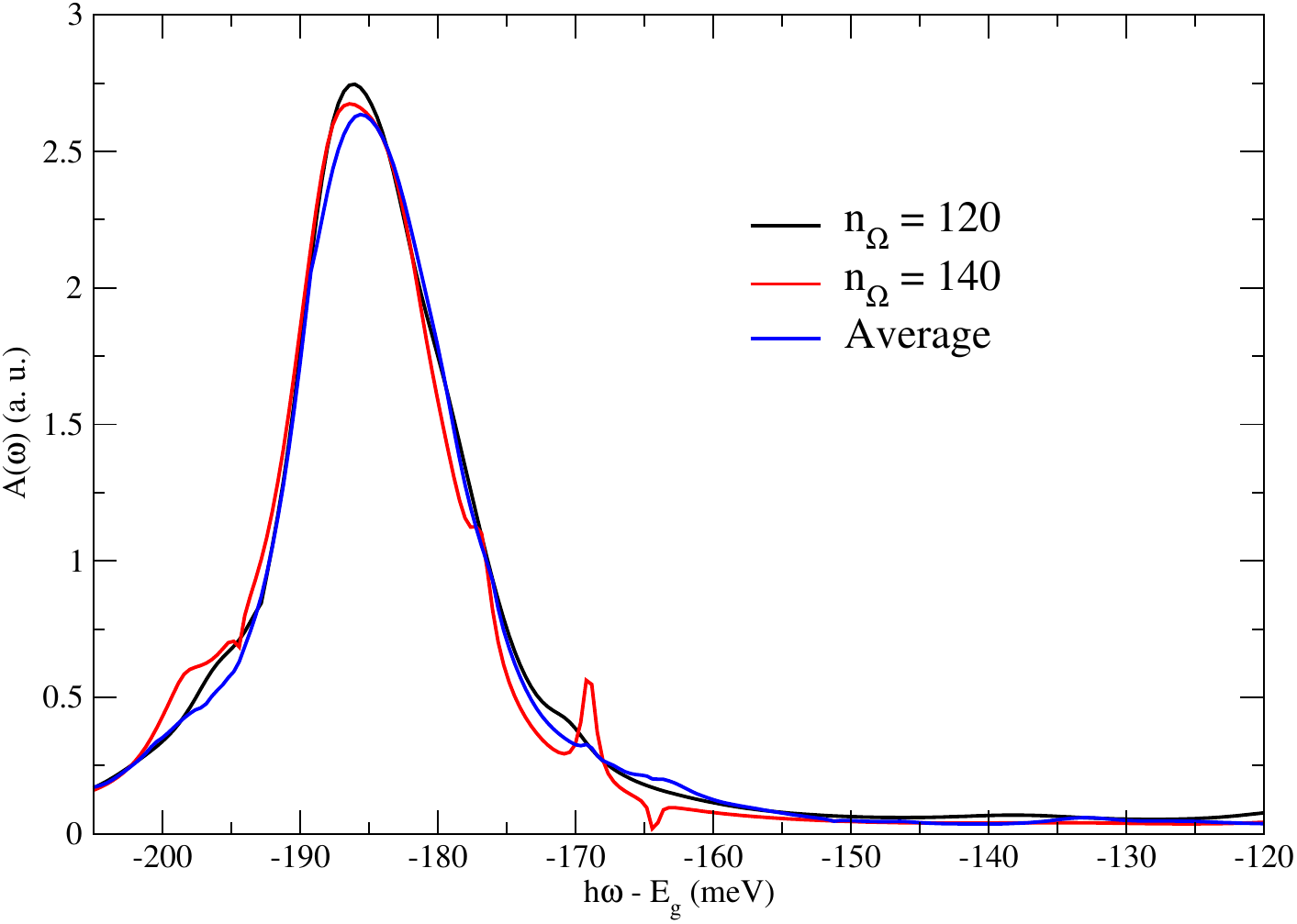}
\caption{ An example of using the analytical continuation to obtain the absorption spectrum of  hBN-WSe$_2$-STO structure. The black and red lines are absorption spectra obtained from the Pad\'{e} approximation using $N \equiv n_\Omega = 120$ and 140 points, respectively. The spurious noise  is suppressed by averaging the absorption spectra of different $ n_\Omega$s (see text).}\label{fig:AvgPade} 
\end{figure}

Figure~\ref{fig:AvgPade} shows the absorption spectra obtained from the Pad\'{e} approximation using $N \equiv n_\Omega= 120$ (black line) and $n_\Omega=140$ (red line) points, corresponding to 40 and 20 neglected Matsubara frequencies around $\Omega = 0$, respectively. Other parameters are provided in Appendix \ref{app:Para}. Because the technique employs the rational fraction in Eq. (\ref{Pade1977}) to approximate the Green's function, it inherently introduces noises in form of random and weak  spurious peaks other than the main broad peak of the exciton bound state. The latter emerges at the same spectral position for  all values of $n_\Omega$. Averaging the spectra of different $n_\Omega$s helps  to suppress the spurious peaks. The results shown in this work are averages  of  spectra calculated with  $n_\Omega = \{120, 122, 124, ...   140\}$.

%%%%%%%%%%%%%%%%%%%%%%%%%%%%%%%%%%%%%%%%%%%%%%%%%%%%%%%%%%%%%%%%%%%%%%%%%%%%%%%
%%%%%%%%%%%%%%%%%%%%%%%%%%%%%%%%%%%%%%%%%%%%%%%%%%%%%%%%%%%%%%%%%%%%%%%%%%%%%%%

\section{Parameters}\label{app:Para}
The following parameters are used for the WSe$_2$ monolayer in different dielectric environments. 
\begin{enumerate}
\item The effective masses are $m_c = 0.29m_0$ (top conduction-band valley), $m_\text{v} = 0.36 m_0$ (top  valence-band valley)  \cite{Kormanyos_2DMater15}. The kinetic energies of electrons and holes are evaluated by parabolic energy dispersion. 
\item The monolayer parameters of the potential are $d = $ 6 $\AA$ and  $\ell_+ = \ell_- = 5 d$ (Appendix \ref{ap:Ponten}).
\item The dielectric  constants of hBN, TiO$_2$, and STO, and the parameter $r_1 = \hbar \omega_\text{TO}/\varepsilon_{\text{X}^0}$ (with $\varepsilon_{\text{X}^0} = 170$ meV)  are listed in  Table \ref{tab:DiePara}. 

\begin{table}[t]
\caption{\label{tab:DiePara} Parameters of different  dielectric environments.}

\begin{center}
\begin{tabular}{||c || c | c | c ||} 
 \hline
 Materials & hBN  & TiO$_2$ & STO \\ [0.5ex] 
 \hline\hline
$\epsilon_0$   & 4.9 \cite{VanTuan_PRB18, Cai_SSC2007, Dai_Science2014}  & 192 \cite{Parker_PR61, Schoche_JAP13} &  25000 \cite{Sawaguchi_JPSJ62, Neville_JAP72}\\ 
 \hline
$\epsilon_\infty$  & 3.8 \cite{VanTuan_PRB18, Cai_SSC2007, Dai_Science2014} & 6 \cite{Schoche_JAP13} & 6 \cite{Neville_JAP72} \\ 
 \hline
$r_1$  & 0.55 \cite{Cai_SSC2007}  & 0.14 \cite{Schoche_JAP13} & 0.1 \cite{Evarestov_PRB2011, Sirenko1999,Akimov_PRL2000} \\ 
 \hline
 \hline
\end{tabular}
\end{center}

\end{table}

\item The following parameters are used in  the numerical calculations of the self-energy and BSE: $n_K=100$, $E^k_\text{Cut} = 2$ eV,  $n_Z = 3200$,  T =  80 K, and the broadening used  in the analytical continuation of Eq. (\ref{Eq:Ana1}) and (\ref{Eq:Ana2}) is $\delta = 3$ meV. 

\end{enumerate}

%%%%%%%%%%%%%%%%%%%%%%%%%%%%%%%%%%%%%%%%%%%%%%%%%%%%%%%%%%%%%%%%%%%%%%%%%%%%%%%
%%%%%%%%%%%%%%%%%%%%%%%%%%%%%%%%%%%%%%%%%%%%%%%%%%%%%%%%%%%%%%%%%%%%%%%%%%%%%%%
\section{Numerical aspects} \label{App:NumAsp}

\begin{figure}[b!] 
\centering
\includegraphics[width=8.5cm]{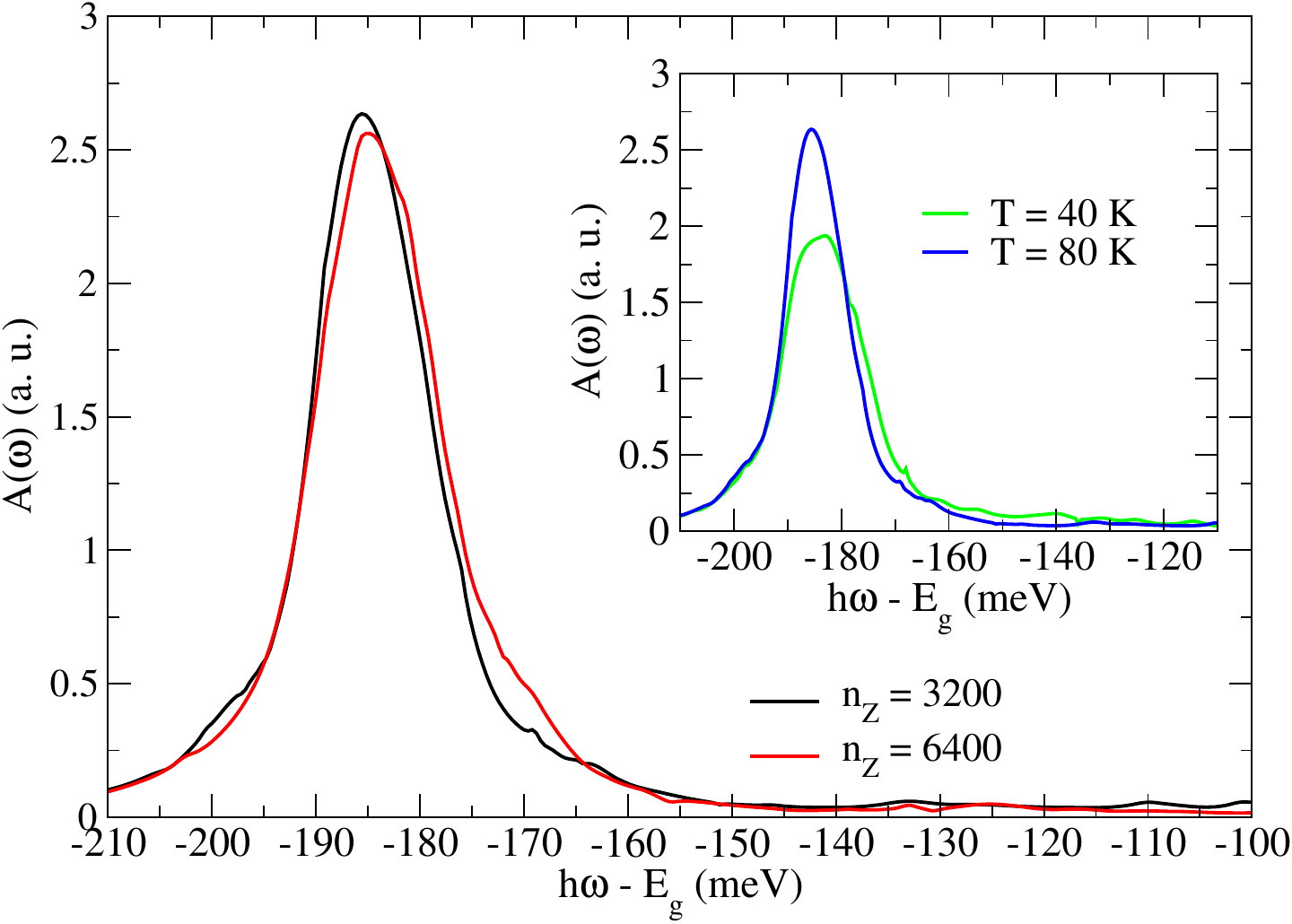}
\caption{Calculated absorption spectra  with $n_Z = 3200$ and $6400$ fermionic Matsubara frequencies. Inset:   Calculated absorption spectra at $T=40$~K and $80$~K.     }\label{fig:TempOmegaDep} 
\end{figure}

The temperature enters  the BSE (Eq. (\ref{Bethe-Salpeter1})) via the imaginary Matsubara frequencies, classified as  bosonic $\Omega = 2 \ell \pi ik_\text{B} T $ (exciton)  or fermionic  $z= (2\ell+1)\pi i k_\text{B}T$ (electron and hole).  Due to  computational limitation, the calculation  includes a finite number of fermionic frequencies $n_Z$ which, as mentioned in Appendix \ref{app:discrete}, is chosen to guarantee a very large $E^z_\text{Cut}$. All the calculations in this work are performed with $n_Z= 3200$ and $T = 80 $ K, corresponding to $E^z_\text{Cut} \simeq 70 $ eV which is far larger than the bandgap and exciton binding energy of the system.   

The converged solutions are shown by the inset of Fig. \ref{fig:TempOmegaDep} in which we compare the results for T = 40 K and T = 80 K.  While doubling the temperature somewhat affects the resonance amplitudes, it hardly changes the resonance energy (we only care for the exciton energy in this work). Doubling the number of   fermionic frequencies, $n_Z$ as shown in Fig. \ref{fig:TempOmegaDep}, hardly affects the height and spectral position of the peak, meaning that the calculation is converged with $T = 80$ K and $n_Z = 3200$.    

It is emphasized  that the temperature $T$ sets the energy resolution of the discrete Matsubara frequencies, while it does not contribute to the broadening of the resonance peak. The temperature-dependent broadening, as seen in experiments, can be modeled through the broadening parameter $\delta(T)$. We have used $\delta =3$~meV throughout this work, neglecting its temperature dependence for simplicity.

%Another numerical aspect is the difficulty to reach convergence at  $\Omega = 0$. Solving the BSE with the iterative method has fast convergence at high bosonic Matsubara frequencies $\Omega$. The convergence is slower for small $\Omega$s and impossible to reach at $\Omega = 0$.  This difficulty  can be explained by the fact that these frequencies are strongly related to states close to the exciton continuum (high-energy states such as $2s, 3s, $ etc.). Each state corresponds to a pole in the exciton Green function.   The closeness of these poles makes  rapid oscillations in the Green  function which is more difficult to capture.   Fortunately, we are interested in the $1s$ state which is well separated from the high-energy states. Therefore, we  do not need the exciton Green function at  low $\Omega$s. Fig. \ref{fig:AvgPade} shows the results  for $n_\Omega = 120$ and $n_\Omega = 140$, corresponding to 40 and 20 neglected Matsubara frequencies around $\Omega = 0$, respectively. The energy positions of the peaks are almost the same for the two cases. 
%

%\bibliographystyle{apsrev}
%\bibliography{BibOptics2D}

%%%%%%%%%%%%%%%%%%%%%%%%%%%%%%%%%%%%%%%%%%%%%%%%%%%%%%%%%%%%%%%%%%%%%%%%%%%%%%%
%%%%%%%%%%%%%%%%%%%%%%%%%%%%%%%%%%%%%%%%%%%%%%%%%%%%%%%%%%%%%%%%%%%%%%%%%%%%%%%
%%%%%%%%%%%%%%%%%%%%%%%%%%%%%%%%%%%%%%%%%%%%%%%%%%%%%%%%%%%%%%%%%%%%%%%%%%%%%%%
%%%%%%%%%%%%%%%%%%%%%%%%%%%%%%%%%%%%%%%%%%%%%%%%%%%%%%%%%%%%%%%%%%%%%%%%%%%%%%%

\end{document}